\documentclass[reprint,aps,prx]{revtex4-2}

\usepackage{amsthm}
\usepackage{amsmath}

\newcommand{\ket}[1]{|#1\rangle}

\newcommand{\inner}[2]{\langle #1|#2\rangle}
\newcommand{\amp}[3]{\langle #1|#2|#3\rangle}
\newcommand{\Hcal}{\mathcal H}
\newcommand{\Ccal}{\mathcal C}
\newcommand{\N}{\mathbf N}
\newcommand{\R}{\mathbf R}
\newcommand{\Thit}{T_{\rm hit}}

\newtheorem{theorem}{Theorem}
\newtheorem{corollary}{Corollary}

\begin{document}

\title{Undecidability of Hitting Times in Computably Described Quantum Dynamics:
A No-Go Theorem for Universal Time Selection}

\author{Katsufumi Matsuura}

\email{2410000373@campus.ouj.ac.jp}
\affiliation{The Open University of Japan, 2-11 Wakaba, Mihama-ku, Chiba 261- 8586 Japan}

\begin{abstract}
We formulate a hitting-time problem for computably described unitary
quantum dynamics. Given computable initial and target states, a
computably described unitary evolution $U(t)$, and a threshold
$0<\varepsilon<1/2$, the unitary hitting-time problem asks for the
first time at which the target fidelity reaches the prescribed
threshold, with value $\infty$ if this never occurs. We prove that no
total algorithm can decide for all such inputs whether the hitting
time is finite, and hence no total algorithm can return the hitting
time universally. More strongly, undecidability persists for a single
fixed computably described unitary evolution generated by a bounded
piecewise-constant Hamiltonian with period $2$; only the initial and
target computational-basis states depend on the encoded Turing
machine and input. Thus the obstruction is not an artifact of
input-dependent time schedules. The result also excludes universal
total time-selection procedures, and persists in a two-sided
formulation allowing $t\in\mathbf R$.
\end{abstract}

\maketitle
\section{Introduction}

In a time-evolving quantum system, it is natural to ask whether a given
initial state reaches a specified target state and, if it does, when this
first occurs. Questions of reachability and characteristic time scales
arise throughout quantum dynamics, including quantum control, quantum
algorithms, recurrence problems, and many-body dynamics.

There is a long-established connection between computation and quantum
dynamics. Classical computation can be made logically reversible
\cite{Bennett1973}, and reversible computation can in turn be represented
within unitary quantum dynamics \cite{Feynman1985,Kitaev2002}. This
connection makes it possible for limitations of classical computability
to appear as limitations on questions formulated directly in terms of
physical dynamics.

Importantly, the argument developed below does not rely on specifically
quantum computational resources such as entanglement, interference, or
many-body structure. At the computational level, the essential
requirement is only that the dynamics admit a computably labelled basis
in which the successive configurations of a reversible computation can
be represented and followed effectively. The obstruction therefore
concerns time as a computational resource: once universal computation
can be embedded into physical evolution, a procedure that must determine
universally whether an event occurs and return the physical time of its
occurrence is subject to the same total-computability obstruction as the
underlying computation.

From this perspective, the relevant unbounded resource is not a
specifically quantum resource such as entanglement or interference,
but the amount of logical evolution that may have to occur before the
target event is reached. This distinguishes the present problem from
undecidability results whose difficulty arises from thermodynamic limits,
spectral structure, control optimization, or specific properties of
quantum measurements.

Algorithmic undecidability has indeed been established in several areas
of quantum theory. The spectral-gap problem provides a prominent
many-body example \cite{Bausch2020}, while uncomputability also appears
in quantum-control problems \cite{Bondar2020}. More directly related to
the present work, Blondel \textit{et al.} studied decidable and
undecidable emptiness problems for quantum finite automata
\cite{Blondel2005}; Eisert \textit{et al.} showed that the occurrence of
outcomes in sequential quantum measurements can be undecidable
\cite{Eisert2012}; and Li and Ying established undecidability for several
reachability properties of quantum systems \cite{LiYing2014}.
Undecidable dynamical behaviour can also persist under strong physical
restrictions: Shiraishi and Matsumoto embedded reversible universal
computation into one-dimensional local Hamiltonian dynamics and showed
that thermalization can be undecidable \cite{Shiraishi2021}.

These results establish that natural yes/no properties of quantum systems
may fail to admit universal decision procedures. The purpose of the
present work is not to establish undecidability of quantum reachability
as such. Instead, we ask what happens when reachability is promoted from
a Boolean decision problem to a total problem whose output is the
physical time associated with the event.

For computably described data
\[
(\ket{\psi},\ket{\phi},U,\varepsilon),
\]
we define the unitary hitting time
\[
T_{\rm hit}
:=
\inf\left\{
t\geq0:
|\langle\phi|U(t)|\psi\rangle|^2\geq1-\varepsilon
\right\},
\]
with
\[
T_{\rm hit}=\infty
\]
when the target is never reached. The unitary hitting-time problem
(UHTP) asks whether there exists a single total procedure which returns
the extended-real-valued quantity
\[
T_{\rm hit}\in[0,\infty]
\]
for every computably described input.

This formulation brings several logically distinct questions into a
single framework. A total solution must first resolve the
\emph{finite/infinite dichotomy}, namely whether the target is reachable
at all. If the target is reachable, it must moreover determine the
\emph{first hitting time}, rather than merely return a Boolean
reachability answer. A total solution would therefore constitute a
\emph{universal time-selection procedure}: for every input it would
either specify a time at which the target is first reached or certify
that no finite target-reaching time exists. When such a procedure is
interpreted operationally, the same question becomes whether a uniformly
specified finite-resource physical protocol can realize this time
selection for all inputs. Finally, the problem admits a two-sided
extension with $t\in\mathbf R$, allowing the same question to be posed
without fixing the sign of the time parameter in advance.

The main result of this paper is that no such universal total procedure
exists.  We prove that no total algorithm can decide, for every
computably described input, whether
\[
T_{\rm hit}<\infty.
\]
Consequently, no total algorithm can return
\[
T_{\rm hit}\in[0,\infty]
\]
for all inputs.

Under the broad definition of computably described dynamics used here,
the finite/infinite reachability alternative already admits a direct
reduction from the halting problem in which the simulated computation
is encoded into a computable time-dependent schedule.  We make this
simple reduction explicit below.

More strongly, we show that such input-dependent time dependence is not
necessary.  There exists a single fixed computably described unitary
evolution $U_*(t)$, independent of the simulated Turing machine and its
input, for which the UHTP remains undecidable when only the initial and
target computational-basis states are allowed to vary.  The fixed
evolution is generated by a bounded piecewise-constant Hamiltonian
$H_*(t)$ that is periodic with period $2$.

For every Turing machine $M$ and input $x$, we effectively construct
computational-basis states
\[
\ket{\psi_{M,x}}
\quad\hbox{and}\quad
\ket{\phi_{M,x}}
\]
such that
\[
M(x)\ {\rm halts}
\quad\Longleftrightarrow\quad
\exists t\geq0:
|\amp{\phi_{M,x}}{U_*(t)}{\psi_{M,x}}|^2
=
1.
\]
Thus the undecidable computation need not be encoded into an
externally supplied time schedule; it can instead be carried by the
state evolving under one and the same fixed dynamical law.

The distinction between reachability undecidability and total
hitting-time computation is central to the present work. Earlier
results already establish undecidability for reachability and related
dynamical predicates
\cite{Blondel2005,Eisert2012,LiYing2014,Shiraishi2021}.
The UHTP instead treats the time of the event itself as the output of a
total problem. In particular, returning $T_{\rm hit}$ requires not only
finding a finite witness when one exists, but also returning $\infty$
when no such witness exists. The reachability decision problem therefore
appears as the finite/infinite boundary of a single time-valued
computational problem.

This viewpoint yields consequences beyond the Boolean reachability
statement. The undecidability of the UHTP implies the nonexistence of a
universal time-selection function. Under the standard effectiveness
assumption that uniformly specified finite-time, finite-resource
physical protocols admit effective finite descriptions, it also gives
an operational no-go result for universal finite-resource time
selection. These conclusions concern universal total procedures and do
not preclude the computation or experimental determination of hitting
times for individual systems or restricted decidable classes.

We further introduce a two-sided UHTP in which $t\in\mathbf R$ is
allowed and show that the same undecidability persists. This extension
does not establish an arrow of time, nor does it rely on a particular
interpretation of time-reversal symmetry. Rather, it shows that allowing
both signs of the time parameter does not restore a universal total
procedure for time selection.

The resulting framework therefore unifies five aspects of the problem:
the finite/infinite reachability dichotomy, total computability of the
first hitting time, universal time selection, finite-resource operational
realization, and two-sided time. The central obstruction concerns not
only whether a target event occurs, but whether the time associated with
that event can be supplied universally as the output of a total
procedure.

The remainder of the paper is organized as follows. Section
\ref{sec:related} reviews related undecidability and reachability results
in more detail. Section \ref{sec:definitions} defines computable quantum
data and the UHTP. Section \ref{sec:main} proves the main undecidability
theorem. Section \ref{sec:consequences} derives the consequences for
universal time selection and finite-resource protocols. Section
\ref{sec:twosided} introduces the two-sided UHTP, and Sections
\ref{sec:discussion} and \ref{sec:conclusion} discuss the scope and
implications of the result.

\section{Background and related work}
\label{sec:related}

The relation between computation and physical dynamics has a long
history. Bennett showed that irreversible classical computation can be
simulated by logically reversible computation \cite{Bennett1973}.
Reversible computational steps can then be represented as permutations
of computational configurations and hence as unitary transformations.
This connection underlies standard constructions of quantum computation
and Hamiltonian implementations of computation
\cite{Feynman1985,Kitaev2002}. It also provides a general mechanism by
which undecidable properties of computation can be transferred to
questions about physical dynamics.

For the present problem, the important point is that this mechanism does
not require specifically quantum computational resources such as
entanglement or interference. The essential structure is a computably
labelled basis in which computational configurations can be represented
and their evolution followed effectively. Once an unbounded
computation can be embedded in physical evolution, the amount of
physical time required before a designated event occurs can encode the
logical running time of that computation. The UHTP is concerned with
the consequences of asking for this physical time as the output of a
total procedure.

Undecidability in quantum theory is already known in several distinct
settings. A prominent example is the spectral-gap problem. Bausch
\textit{et al.} showed that spectral-gap undecidability persists even for
one-dimensional quantum spin systems \cite{Bausch2020}. Here the
undecidable object is a spectral property associated with a family of
many-body Hamiltonians and its thermodynamic-limit behaviour. This
provides an important example of undecidability intrinsic to a
physically constrained quantum model, but the question is not formulated
as the time at which a prescribed dynamical event occurs.

A different class of problems arises in quantum control. Bondar and
Pechen studied uncomputability and computational complexity in the
control of quantum systems \cite{Bondar2020}. Such problems ask, for
example, whether suitable controls exist or whether an appropriate
control can be computed. They are therefore closely related to
reachability, but the control variables themselves form part of the
optimization or synthesis problem. In the UHTP, by contrast, the
dynamics is specified and the quantity sought is the time associated
with reaching a prescribed target.

More directly related results concern reachability and occurrence
problems. Blondel \textit{et al.} studied emptiness and threshold
problems for quantum finite automata and established both decidable and
undecidable cases \cite{Blondel2005}. Their setting asks whether there
exists an input word whose acceptance probability satisfies a specified
condition. The central output is therefore a Boolean existence
property rather than the time of the corresponding event.

Eisert \textit{et al.} considered repeated quantum measurements and
showed that the occurrence of measurement outcomes can be undecidable
\cite{Eisert2012}. Their result demonstrates that even for
finite-dimensional quantum systems it may be impossible to decide
universally whether certain finite sequences of measurement outcomes can
occur. This is particularly relevant to the present work because it
shows that undecidability may arise in an operationally formulated
quantum problem without relying on a thermodynamic limit. Nevertheless,
the question remains an occurrence problem: whether a specified event
can happen.

Li and Ying studied reachability of quantum systems more explicitly
\cite{LiYing2014}. They distinguished several temporal properties,
including whether a target subspace is eventually reached, always
reached, eventually always reached, or reached infinitely often, and
showed undecidability for the corresponding general problems. These
results establish that undecidability of quantum reachability itself is
not new. The contribution of the present work should therefore not be
understood as the observation that a target-state reachability predicate
can encode an undecidable problem.

Another closely related direction embeds universal computation directly
into Hamiltonian dynamics. Shiraishi and Matsumoto constructed a
one-dimensional local Hamiltonian system in which reversible universal
computation is embedded into the dynamics and used this construction to
show undecidability of thermalization \cite{Shiraishi2021}. This result
is especially relevant because it demonstrates that undecidable
dynamical properties can persist under strong restrictions on the
Hamiltonian. The property considered there, however, is the asymptotic
thermalization behaviour of the system rather than the total
computability of the time of a specified event.

Against this background, the UHTP differs from these earlier
reachability and occurrence problems by treating the time of the event
itself as the output of a total problem.  For computably described
initial and target states and a computably described unitary evolution,
we define
\[
T_{\rm hit}
:=
\inf\left\{
t\geq0:
|\langle\phi|U(t)|\psi\rangle|^2\geq1-\varepsilon
\right\},
\]
with $T_{\rm hit}=\infty$ when the target is never reached.  Thus the
output space of the problem is
\[
[0,\infty],
\]
rather than a Boolean set.

This change of output exposes a distinction that is hidden in a pure
reachability formulation.  A total UHTP solver must simultaneously
perform two tasks.  It must determine whether the target is reachable,
that is, distinguish
\[
T_{\rm hit}<\infty
\qquad\hbox{from}\qquad
T_{\rm hit}=\infty,
\]
and, in the finite case, determine the first time at which the prescribed
threshold is reached.  The finite/infinite reachability decision is
therefore contained as the boundary between the two classes of outputs,
rather than being the complete problem.

This time-valued formulation also gives the problem an immediate
operational interpretation.  A total solver would constitute a universal
time-selection procedure: given the dynamical description and a target,
it would specify how far the evolution must proceed in order to reach
that target, or return $\infty$ when no such time exists.  If the solver
is required to be physically realized, the same formulation leads to
the question of whether such time selection can be performed by a
uniformly specified finite-resource protocol.  These consequences are
specific to treating the hitting time itself, rather than only
reachability, as the object to be computed.

The distinction between the broad input class and fixed dynamics is
also important.  Under the general definition of computably described
dynamics used in this paper, reachability undecidability already admits
a direct reduction in which the simulated computation is encoded into
an input-dependent time schedule.  The stronger result proved below
shows that this freedom is not essential.  There exists a single fixed
computably described unitary evolution $U_*(t)$, independent of the
simulated Turing machine and its input, for which the UHTP remains
undecidable when only the initial and target computational-basis states
are allowed to vary.  Thus the computability obstruction persists even
when the dynamical law itself is held fixed.

Finally, the same framework admits a two-sided extension in which
$t\in\mathbf R$.  This makes it possible to ask for universal time
selection without fixing the sign of the time parameter in advance.  The
two-sided formulation is not intended as a derivation of a physical
arrow of time; its role is to test whether allowing both temporal
directions removes the computability obstruction.  As shown below, it
does not.

The present work therefore builds on, rather than replaces, earlier
undecidability results for reachability, measurement occurrence,
control, spectral properties, and dynamical behaviour.  Its distinct
focus is the total computation of time itself.  Within a single
framework, the UHTP connects the finite/infinite reachability dichotomy,
the computability of the first hitting time, universal time selection,
finite-resource operational realization, and the corresponding
two-sided problem.  The main theorem further shows that this obstruction
already persists under a single fixed computably described unitary
evolution.
Thus the combined time-valued problem admits no universal total
solution.

\section{Computable quantum dynamics and the UHTP}
\label{sec:definitions}

\subsection{Computable states}

Let $\Hcal$ be a separable Hilbert space with a fixed computably
labelled orthonormal basis
\[
\{\ket{c}: c\in\Ccal\},
\]
where $\Ccal$ is a countable set of finite strings or finite classical
configurations.

A pure state $\ket{\psi}\in\Hcal$ is called computable if, for every
precision $2^{-m}$, there exists an algorithm which outputs a finite
rational complex linear combination
\[
\ket{\psi_m}
 =
 \sum_{j=1}^{N_m} a_j\ket{c_j}
\]
such that
\[
\|\ket{\psi}-\ket{\psi_m}\|\leq 2^{-m}.
\]
In the reduction used below, the initial state and the target state
are both single computational basis states\cite{Feynman1985,Kitaev2002}.  Thus only the simplest
case of this general definition is used.

\subsection{Computably described unitary evolution}

A unitary evolution $U(t)$, $t\geq0$, with $U(0)=I$, is called computably described
if, for every computable finite-support state $\ket{\xi}$, rational
time $t\geq0$, and precision $2^{-m}$, there is an algorithm which
outputs a finite description of a $2^{-m}$-approximation to
$U(t)\ket{\xi}$.  We additionally require that, for every computable finite-support state
$\ket{\xi}$ and every finite time interval $[0,T]$, the map
\[
t\longmapsto U(t)\ket{\xi}
\]
has a computable modulus of continuity on $[0,T]$.  The approximation procedure and the modulus of continuity are required
to be uniform in the effective descriptions of their arguments.

This definition specifies the broad effective input class of the UHTP.
It does not require the time dependence of the evolution to be
independent of the particular problem instance.  In particular, it allows a bounded piecewise-constant Hamiltonian with
a computable uniform norm bound, whose value on the $n$th time
interval is determined by a finite computation depending on the input.
If the switching times are effectively computable and the rule
determining each interval terminates after finitely many computational
steps, then the resulting schedule is computably described in the
above sense.

This freedom makes the broad input class sufficiently general that
reachability undecidability already admits a direct reduction from the
halting problem.  As shown explicitly in Section \ref{sec:main}, one
may encode the simulation of a Turing machine into the algorithm
generating a sequence of identity and $X$ operations on a single qubit.
Such a construction establishes undecidability for the broad class
defined here, but places the simulated computation in the externally
specified time dependence.

The main result proved below is stronger and does not rely on this
freedom.  We construct a single fixed computably described unitary
evolution $U_*(t)$, independent of the simulated Turing machine $M$ and
its input $x$, such that the UHTP remains undecidable when only the
initial and target computational-basis states are allowed to vary.
The corresponding Hamiltonian $H_*(t)$ is bounded, piecewise constant,
and periodic with period $2$.  Thus the undecidable computation is
carried by the state evolving under a fixed dynamical law rather than
by an input-dependent schedule.

More generally, a bounded piecewise-constant Hamiltonian with a
computable uniform bound on $\|H(t)\|$, and with effectively computable
switching times and action on computational basis states, generates a
computably described evolution
\[
U(t)
=
{\mathcal T}
\exp\left(
-i\int_0^t H(s)\,ds
\right).
\]

For the fixed evolution used in the main theorem, the action of each
Hamiltonian segment can in fact be evaluated explicitly.  Consequently,
for every rational $t\geq0$, every computable finite-support state
$\ket{\xi}$, and every precision $2^{-m}$, the state
\[
U_*(t)\ket{\xi}
\]
can be approximated effectively to the required precision.

The definition above therefore contains the natural broad class of
effectively specified quantum dynamics for which the UHTP is
formulated, while the stronger fixed-dynamics theorem shows that the
undecidability result is not an artifact of allowing the
time-dependent schedule itself to encode the input computation.

\subsection{The unitary hitting-time problem}

Fix $0<\varepsilon<1/2$.  For a computable initial state $\ket{\psi}$,
a computable target state $\ket{\phi}$, and a computably described
unitary evolution $U(t)$, define
\[
\Thit(\psi,\phi,U,\varepsilon)
 :=
 \inf\left\{t\geq 0:
 |\amp{\phi}{U(t)}{\psi}|^2\geq 1-\varepsilon
 \right\}.
\]
If the set is empty, we set
\[
\Thit(\psi,\phi,U,\varepsilon)=\infty.
\]

We call this problem the unitary hitting-time problem, or UHTP.  The
UHTP asks for a procedure which, given
\[
(\ket{\psi},\ket{\phi},U,\varepsilon),
\]
returns the extended nonnegative real value
\[
\Thit\in[0,\infty].
\]

\subsection{Total and partial procedures}

The crucial point is totality.

For the reduction instances constructed below, a partial procedure is
straightforward.  One may evaluate the fixed dynamics at successive
even integer times
\[
t=0,2,4,\ldots
\]
and halt once the target condition is satisfied.  Such a procedure
eventually halts whenever the encoded computation halts, but continues
indefinitely otherwise.

The problem studied in this paper is the existence of a total
procedure: a procedure which halts on every input, returns a finite
time when the target is reached, and returns $\infty$ when the target
is unreachable.

This totality requirement is exactly where the connection with the
halting problem appears.  In general, it is impossible to certify in
finite time, uniformly for all inputs, that the target will never be
reached.

\section{Undecidability of the UHTP}
\label{sec:main}

We now prove the main undecidability result. We first point out that,
under the broad definition of computably described dynamics introduced
in Section \ref{sec:definitions}, reachability undecidability admits a
very simple construction. We then give a stronger construction showing
that the undecidability does not rely on encoding the input computation
into an externally supplied time-dependent schedule.

\subsection{A direct reduction using a computable time-dependent schedule}

Under the general input model of Section \ref{sec:definitions}, the
halting problem can be encoded directly into the time dependence of a
single-qubit evolution.

Let $M$ be an arbitrary Turing machine and let $x$ be an arbitrary
input. Consider a qubit initially in the state $\ket{0}$.
For each positive integer $n$, determine the gate applied during the
$n$th time interval by simulating $M(x)$ for $n$ computational steps.
If $M(x)$ halts for the first time at step $n$, apply an $X$ gate during
that interval; otherwise apply the identity.

This gate schedule is computable. Determining the gate used at the
$n$th step requires only a finite simulation of $M(x)$ for $n$ steps
and therefore does not require deciding whether $M(x)$ ever halts.

If $M(x)$ never halts, the state remains $\ket{0}$ for all times. If
$M(x)$ halts after a finite number of steps, an $X$ gate is eventually
applied and the state reaches $\ket{1}$. Hence
\[
M(x)\ {\rm halts}
\quad\Longleftrightarrow\quad
\ket{1}\ {\rm is\ reached}.
\]

A computable sequence of identity and $X$ gates can be implemented by a
computably described bounded piecewise-constant Hamiltonian of the type
allowed in Section \ref{sec:definitions}. Thus this simple construction
already establishes undecidability of the finite/infinite reachability
alternative for the broad class of computably described
time-dependent dynamics.

However, this construction places the simulation of $M(x)$ in the
algorithm generating the external time dependence. It therefore does
not show that the same obstruction persists when the dynamical law is
fixed independently of the simulated Turing machine and its input.

The remainder of this section establishes this stronger statement.

\subsection{Fixed universal reversible computation}

Let $R_*$ be a fixed universal reversible Turing machine. By standard
reversible-computation constructions, an arbitrary Turing machine $M$
and input $x$ can be encoded into a finite initial configuration
\[
c_0(M,x)
\]
such that $R_*$ reversibly simulates the computation of $M(x)$
\cite{Bennett1973}.

The transition rules of $R_*$ are fixed once and for all. They do not
depend on $M$, on $x$, or on the running time of the simulated
computation. The description of $(M,x)$ appears only in the initial
configuration.

The reversible simulator is arranged so that, if the simulated
computation halts, it subsequently reverses the auxiliary computation
history and reaches a canonical marked configuration
\[
c_B(M,x).
\]
The configuration $c_B(M,x)$ retains the encoded input data together
with a dedicated beacon marker, while the auxiliary work registers are
returned to a standard form.

If the simulated computation never halts, this marked configuration is
never reached.

Consequently, there exists a single fixed computable reversible
bijection
\[
w_*:\Ccal\to\Ccal
\]
such that, for every pair $(M,x)$,
\[
\exists n\in\N:
w_*^n\bigl(c_0(M,x)\bigr)
=
c_B(M,x)
\quad\Longleftrightarrow\quad
M(x)\ {\rm halts}.
\]

Both $c_0(M,x)$ and $c_B(M,x)$ are effectively computable from
$(M,x)$. Each is a single classical configuration and therefore
corresponds to a single computational basis state.

In the present construction, retaining the encoded input in
$c_B(M,x)$ ensures that the reversible map remains one-to-one when
the auxiliary computation history is uncomputed. Distinct reversible computation histories cannot merge
into a single configuration under a one-to-one dynamical map. The
important point is that the dynamical law $w_*$ itself is fixed; only
the initial and target states depend on the UHTP input.

\subsection{Fixed permutation unitary}

The fixed reversible map $w_*$ defines a single permutation unitary
$W_*$ on $\ell^2(\Ccal)$ by
\[
W_*\ket{c}
=
\ket{w_*(c)}.
\]

Since $w_*$ is fixed independently of $(M,x)$, the unitary $W_*$ is
also fixed once and for all.  The dependence on the simulated Turing
machine and its input appears only through the computational basis
states corresponding to the initial and beacon configurations.

By construction,
\[
\exists n\in\N:
W_*^n\ket{c_0(M,x)}
=
\ket{c_B(M,x)}
\quad\Longleftrightarrow\quad
M(x)\ {\rm halts}.
\]

Thus the undecidable computation is not encoded into a
machine-dependent sequence of unitary operators.  Rather, the same
fixed permutation unitary $W_*$ acts on different computational basis
states encoding different pairs $(M,x)$.

At this stage the construction is still discrete in time.  We next
embed this single fixed reversible dynamics into a fixed
continuous-time unitary evolution.  The initial and target states of
the UHTP reduction will be defined after introducing the auxiliary
two-level degree of freedom required for that embedding.

\subsection{Embedding into a fixed continuous-time unitary evolution}

The embedding of reversible universal computation into quantum
Hamiltonian dynamics is well established. In particular, Shiraishi and
Matsumoto constructed quantum many-body Hamiltonians encoding the
dynamics of a reversible universal Turing machine, and showed that such
computational embeddings can be realized even in one-dimensional
shift-invariant systems with nearest-neighbour interactions
\cite{Shiraishi2021}. Their construction is used for a different
purpose, namely the undecidability of thermalization. For the present
hitting-time problem, we require only a considerably simpler
continuous-time realization of the fixed reversible map $w_*$.

Introduce one auxiliary two-level degree of freedom and consider the
extended Hilbert space
\[
\widetilde{\Hcal}
=
\ell^2(\Ccal)\otimes\mathbf{C}^2,
\]
with computational basis states denoted by
\[
\ket{c,b},
\qquad
c\in\Ccal,\quad b\in\{0,1\}.
\]

Define two operators $A$ and $B$ on this basis.  The first operator
flips the auxiliary degree of freedom:
\[
A\ket{c,0}
=
\ket{c,1},
\qquad
A\ket{c,1}
=
\ket{c,0}.
\]

The second operator implements one step of the reversible computation
while flipping the auxiliary degree of freedom:
\[
B\ket{c,1}
=
\ket{w_*(c),0},
\qquad
B\ket{c,0}
=
\ket{w_*^{-1}(c),1}.
\]

Since $w_*$ is a reversible bijection, both $A$ and $B$ are
well-defined permutation unitaries.  Moreover,
\[
A^2=B^2=I,
\qquad
A=A^\dagger,
\qquad
B=B^\dagger.
\]

Their product implements exactly one step of the fixed reversible
computation on the sector with auxiliary state $\ket{0}$:
\[
BA\ket{c,0}
=
\ket{w_*(c),0}.
\]

We now realize $A$ and $B$ as finite-time unitary evolutions.  Define
the bounded self-adjoint Hamiltonians
\[
H_A
=
\frac{\pi}{2}(I-A)
\]
and
\[
H_B
=
\frac{\pi}{2}(I-B).
\]

Because $A$ and $B$ are self-adjoint involutions,
\[
e^{-iH_A}
=
A,
\qquad
e^{-iH_B}
=
B.
\]

Define a single time-dependent Hamiltonian $H_*(t)$ by repeating these
two Hamiltonians with unit duration:
\[
H_*(t)
=
\begin{cases}
H_A,
&
2n\leq t<2n+1,
\\
H_B,
&
2n+1\leq t<2n+2,
\end{cases}
\qquad
n=0,1,2,\ldots.
\]

Let
\[
U_*(t)
=
{\mathcal T}
\exp\left(
-i\int_0^t H_*(s)\,ds
\right)
\]
be the corresponding unitary evolution.

Since $w_*$ and $w_*^{-1}$ are computable on finite configurations,
the actions of $A$ and $B$ on computational basis states are
computable.  Since $A^2=B^2=I$, their finite-time exponentials are
computable explicitly, and hence $U_*(t)\ket{\xi}$ can be approximated
effectively to arbitrary precision for every rational $t$ and every
computable finite-support state $\ket{\xi}$.

Moreover,
\[
\|H_*(t)\|\leq\pi
\qquad
{\rm for\ all}\quad t\geq0.
\]
Hence, for every state $\ket{\xi}$,
\[
\|U_*(t)\ket{\xi}-U_*(s)\ket{\xi}\|
\leq
\pi |t-s|\,\|\ket{\xi}\|.
\]
Thus the fixed evolution has an explicit computable modulus of
continuity and satisfies the computability requirements of
Section \ref{sec:definitions}.

Since the first unit interval implements $A$ and the second implements
$B$, every pair of intervals implements one step of the reversible
computation.  Thus, at every even integer time,
\[
U_*(2n)\ket{c,0}
=
(BA)^n\ket{c,0}
=
\ket{w_*^n(c),0}.
\]

For the reduction associated with $(M,x)$, define the initial state by
\[
\ket{\psi_{M,x}}
:=
\ket{c_0(M,x),0}
\]
and the target state by
\[
\ket{\phi_{M,x}}
:=
\ket{c_B(M,x),0}.
\]

Suppose first that $M(x)$ halts.  Then there exists a finite
computational step $n_B$ such that
\[
w_*^{n_B}\bigl(c_0(M,x)\bigr)
=
c_B(M,x).
\]
Consequently,
\[
U_*(2n_B)\ket{\psi_{M,x}}
=
\ket{\phi_{M,x}}.
\]
Therefore
\[
|\amp{\phi_{M,x}}{U_*(2n_B)}{\psi_{M,x}}|^2
=
1.
\]

Conversely, suppose that $M(x)$ does not halt. At the beginning of
each interval generated by $H_A$, the state is a computational basis
state of the form
\[
\ket{w_*^n(c_0(M,x)),0}.
\]
During that interval, $H_A$ mixes only the two states
\[
\ket{w_*^n(c_0(M,x)),0}
\quad\hbox{and}\quad
\ket{w_*^n(c_0(M,x)),1}.
\]
During the following interval, $H_B$ mixes only
\[
\ket{w_*^n(c_0(M,x)),1}
\quad\hbox{and}\quad
\ket{w_*^{n+1}(c_0(M,x)),0}.
\]

Therefore the target state
$\ket{\phi_{M,x}}=\ket{c_B(M,x),0}$
can acquire nonzero amplitude only if the reversible computational orbit
has reached $c_B(M,x)$ or is about to reach it in the next computational
step.  Either possibility implies that $M(x)$ halts.

Hence, if $M(x)$ does not halt,
\[
|\amp{\phi_{M,x}}{U_*(t)}{\psi_{M,x}}|^2
=
0
\qquad
{\rm for\ all}\quad t\geq0.
\]

Combining the halting and nonhalting cases, we obtain
\[
M(x)\ {\rm halts}
\quad\Longleftrightarrow\quad
\exists t\geq0:
|\amp{\phi_{M,x}}{U_*(t)}{\psi_{M,x}}|^2
=
1.
\]

We have therefore constructed a single fixed periodic bounded
Hamiltonian schedule $H_*(t)$ for which target-state reachability
encodes the halting problem.  The simulated Turing machine and its input
occur only in the initial and target computational-basis states.  In
contrast to the direct single-qubit construction of Section 4.1, no
part of the externally specified time dependence simulates $M(x)$.
The unbounded computation is performed entirely by the state evolving
under the same fixed periodic dynamics.

\subsection{Main theorem}

\begin{theorem}[Undecidability of the UHTP under fixed dynamics]
\label{thm:fixed-uhtp}
There exists a single fixed computably described unitary evolution
$U_*(t)$ such that, for every fixed $0<\varepsilon<1/2$, there is no
total algorithm which, for every pair of computational-basis states
\[
(\ket{\psi},\ket{\phi}),
\]
decides whether
\[
\Thit(\psi,\phi,U_*,\varepsilon)<\infty.
\]

Consequently, even when the unitary evolution $U_*(t)$ is fixed once and
for all, there is no total algorithm which returns
\[
\Thit(\psi,\phi,U_*,\varepsilon)\in[0,\infty]
\]
for every such pair of initial and target states.
\end{theorem}

\begin{proof}
Let $M$ be an arbitrary Turing machine and let $x$ be an arbitrary
input.  From $(M,x)$ we can effectively construct the computational
basis states
\[
\ket{\psi_{M,x}}
=
\ket{c_0(M,x),0}
\]
and
\[
\ket{\phi_{M,x}}
=
\ket{c_B(M,x),0}.
\]

The unitary evolution $U_*(t)$ is fixed and does not depend on $M$ or
$x$.

First suppose that $M(x)$ halts.  By the construction above, there
exists a finite computational step $n_B$ such that
\[
w_*^{n_B}\bigl(c_0(M,x)\bigr)
=
c_B(M,x).
\]
By the continuous-time embedding,
\[
U_*(2n_B)\ket{\psi_{M,x}}
=
\ket{\phi_{M,x}}.
\]
Therefore
\[
|\amp{\phi_{M,x}}{U_*(2n_B)}{\psi_{M,x}}|^2
=
1.
\]
Hence, for every fixed $0<\varepsilon<1/2$,
\[
\Thit
(\psi_{M,x},\phi_{M,x},U_*,\varepsilon)
<
\infty.
\]

Conversely, suppose that $M(x)$ does not halt.  Then the reversible
computation starting from $c_0(M,x)$ never reaches the beacon
configuration $c_B(M,x)$.  By construction, the continuous-time
evolution remains orthogonal to the corresponding beacon state for all
times.  Hence
\[
|\amp{\phi_{M,x}}{U_*(t)}{\psi_{M,x}}|^2
=
0
\qquad
{\rm for\ all}\quad t\geq0.
\]
Therefore
\[
\Thit
(\psi_{M,x},\phi_{M,x},U_*,\varepsilon)
=
\infty.
\]

We have thus effectively constructed, from $(M,x)$, a pair of
computational-basis states such that
\[
M(x)\ {\rm halts}
\quad\Longleftrightarrow\quad
\Thit
(\psi_{M,x},\phi_{M,x},U_*,\varepsilon)
<
\infty.
\]

If a total algorithm deciding whether $\Thit<\infty$ existed for all
pairs of initial and target states under the fixed dynamics $U_*(t)$,
then applying it to
\[
(\ket{\psi_{M,x}},\ket{\phi_{M,x}})
\]
would decide whether $M(x)$ halts.  This contradicts the undecidability
of the halting problem.  Consequently no such total decision algorithm
exists.

Finally, any total algorithm returning the complete value
\[
\Thit\in[0,\infty]
\]
would in particular decide whether that value is finite or infinite.
Hence no such total hitting-time algorithm exists.
\end{proof}

\begin{corollary}
The UHTP is undecidable for the full class of computably described
unitary evolutions defined in Section \ref{sec:definitions}.
\end{corollary}

\begin{proof}
The fixed evolution $U_*(t)$ constructed above is a member of that
class. A total solver for the full class would therefore also solve the
restricted fixed-dynamics problem of Theorem
\ref{thm:fixed-uhtp}, which is impossible.
\end{proof}

\subsection{Role of the reversible-computation construction}

The distinction between the direct single-qubit reduction and Theorem
\ref{thm:fixed-uhtp} is important.

In the direct reduction of Section 4.1, the algorithm generating the
time-dependent schedule simulates $M(x)$ for progressively increasing
numbers of computational steps. The external time dependence therefore
carries the computation whose halting behaviour is being encoded.

By contrast, in Theorem \ref{thm:fixed-uhtp}, the Hamiltonian
$H_*(t)$ and unitary evolution $U_*(t)$ are fixed independently of
$(M,x)$. The input computation is encoded only in the computational
basis state from which the fixed dynamics begins, together with the
corresponding target state.

The reversible-computation construction therefore provides content
beyond the direct reduction available under the broad input definition.
It shows that undecidability of the UHTP is not an artifact of allowing
an input-dependent computable time schedule. The obstruction persists
when the dynamical law itself is held fixed.

In this stronger construction, the physical evolution performs the
unbounded logical computation internally. If the simulated computation
halts, a target event occurs after a finite physical time; if it does
not halt, no such finite time exists. The elapsed physical time thus
carries the logical running time of the embedded computation.

This is the sense in which time functions as the relevant computational
resource in the UHTP.

\subsection{Role of totality}

The theorem does not contradict the existence of partial procedures
which find a finite hitting time when the target is reached.

For example, one may simulate the fixed dynamics $U_*(t)$ for
successively increasing times and halt when the fidelity with the target
state exceeds the prescribed threshold. For the reduction instances
constructed above, such a procedure eventually halts whenever $M(x)$
halts.

If $M(x)$ does not halt, however, the target state is never reached and
the procedure continues indefinitely. Such a procedure therefore
cannot return $\infty$.

The undecidability lies precisely in the requirement of totality. A
universal UHTP solver must not only return a finite witness when one
exists; it must also terminate and return $\infty$ when no finite
hitting time exists. Theorem \ref{thm:fixed-uhtp} shows that this
finite/infinite certification is impossible even when the physical
dynamics itself is fixed.

\section{Consequences: universal time selection and finite-resource protocols}
\label{sec:consequences}

\subsection{Nonexistence of universal time selection}

The undecidability of the UHTP implies the nonexistence of a universal
procedure for selecting a time at which the target is reached.

\begin{corollary}[Nonexistence of universal time selection]
Let $0<\varepsilon<1/2$.  There is no total computable function $S$
which, for every computably described input
\[
(\ket{\psi},\ket{\phi},U,\varepsilon),
\]
satisfies the following conditions:
\begin{enumerate}
\item If the target is reachable, then $S<\infty$ and
\[
|\amp{\phi}{U(S)}{\psi}|^2\geq 1-\varepsilon.
\]
\item If the target is unreachable, then $S=\infty$.
\end{enumerate}
\end{corollary}

\begin{proof}
If such an $S$ existed, then on the UHTP inputs constructed from the
halting problem one could decide whether $M(x)$ halts simply by
checking whether $S$ is finite or infinite.  This contradicts the
undecidability of the halting problem.
\end{proof}

The same conclusion already holds under the single fixed evolution
$U_*(t)$ constructed in Section \ref{sec:main}.  In particular, there
is no total computable function which, for every pair of
computational-basis states $(\ket{\psi},\ket{\phi})$, returns a finite
target-reaching time when one exists under $U_*(t)$ and returns
$\infty$ otherwise.  Thus the failure of universal time selection does
not arise from having to analyze an arbitrary input-dependent
dynamical law; it persists even when the physical evolution is fixed
once and for all.

This corollary shows that undecidability persists even for a weaker
task than returning the first hitting time.  It is not necessary to
return the earliest hitting time.  Even returning some witness time
when the target is reachable and returning $\infty$ when it is not is
impossible in general.

\subsection{Operational consequence for finite-resource protocols}

We next state an operational consequence of the mathematical
undecidability result.

By a physical protocol we mean a procedure consisting of preparation
of the input, implementation of the time evolution, measurement, and
classical post-processing. If such a protocol is specified uniformly
by a finite description of the apparatus, control rules, and
post-processing steps, then it may be regarded as having an effective
finite description.

\begin{corollary}[No uniformly bounded universal physical protocol]
Consider a uniform physical protocol with an effective finite
description. Suppose that, for every computably described input, the
protocol correctly determines whether
\[
\Thit(\psi,\phi,U,\varepsilon)<\infty
\]
or
\[
\Thit(\psi,\phi,U,\varepsilon)=\infty
\]
and, whenever the hitting time is finite, returns the value of
$\Thit$ to a prescribed finite precision. Suppose further that its
observation time and required resources are bounded by
input-independent finite constants. No such universal protocol
exists.
\end{corollary}

\begin{proof}
If such a protocol existed, then applying it to the inputs constructed
from the halting problem would determine whether the corresponding
hitting time is finite or infinite, and hence whether $M(x)$ halts.
Since the protocol is assumed to terminate using uniformly bounded
time and resources, this would give a total decision procedure for the
halting problem. This is impossible.
\end{proof}

The fixed-dynamics construction strengthens this operational
consequence.  For the reduction instances of Theorem
\ref{thm:fixed-uhtp}, the protocol need not be capable of implementing
an arbitrary input-dependent evolution.  The same fixed Hamiltonian
schedule $H_*(t)$ is used for every instance; only the preparation of
the initial computational-basis state and the specification of the
target basis state depend on $(M,x)$.  Hence a uniformly bounded
protocol capable of solving the hitting-time problem even for this
single fixed dynamics would already decide the halting problem.

This consequence does not claim that finite-time experiments or
numerical computations are meaningless for individual physical
systems. What is excluded is a universal protocol which works for all
computably described unitary quantum dynamics and correctly
distinguishes finite hitting times from unreachable targets under
input-independent finite resource bounds.

Nor does the argument derive a detailed lower bound on thermodynamic
work or dissipation in a particular implementation. The claim is more
basic: if the finite/infinite alternative for $\Thit$ could be
decided universally with uniformly bounded finite resources, then the
halting problem could be solved.

\section{Two-sided UHTP and formulations without a fixed time orientation}
\label{sec:twosided}

\subsection{Definition of the two-sided problem}

The UHTP above was formulated for $t\geq 0$.  In foundational physics,
however, it is often important to consider formulations in which the
orientation or choice of a time parameter is not fixed in advance.  We
therefore introduce a two-sided version of the hitting-time problem in
which both positive and negative values of the time parameter are
allowed.

The result proved here does not construct a particular
time-reversal-symmetric dynamical system.  The claim is more limited:
even if the hitting-time problem is extended to $t\in\mathbf R$, the
nonexistence of a universal total solution persists.

Let $U(t)$, $t\in\R$, be a computably described unitary evolution for
all real times.  Define the two-sided hitting time by
\[
T_{\rm hit}^{\pm}(\psi,\phi,U,\varepsilon)
 :=
 \inf\left\{|t|:
 t\in\R,\ 
 |\amp{\phi}{U(t)}{\psi}|^2\geq 1-\varepsilon
 \right\}.
\]
If the set is empty, we set $T_{\rm hit}^{\pm}=\infty$.

\subsection{Undecidability of the two-sided UHTP}

\begin{theorem}[Undecidability of the two-sided UHTP]
There is no total algorithm which decides, for every computably
described two-sided input
\[
(\ket{\psi},\ket{\phi},U,\varepsilon),
\]
whether
\[
T_{\rm hit}^{\pm}(\psi,\phi,U,\varepsilon)<\infty.
\]
Consequently, there is no total algorithm which returns
$T_{\rm hit}^{\pm}$ for all computably described two-sided inputs.
\end{theorem}

\begin{proof}
We reduce the one-sided UHTP to the two-sided UHTP.

Let
\[
(\ket{\psi},\ket{\phi},U,\varepsilon)
\]
be a one-sided UHTP input.  Define a two-sided unitary evolution
$\widetilde U(t)$ by
\[
\widetilde U(t)=
\begin{cases}
U(t),& t\geq 0,\\
I, & t<0.
\end{cases}
\]
For formulations requiring stronger regularity near $t=0$, the
negative-time side may be replaced by any computable interpolation
which keeps the evolved initial state away from the target state.  The
reduction is unaffected.

For $t<0$,
\[
\widetilde U(t)=I,
\]
so the target fidelity is
\[
|\amp{\phi}{\widetilde U(t)}{\psi}|^2
=
|\inner{\phi}{\psi}|^2.
\]
This is exactly the target fidelity at $t=0$ in the original one-sided
problem.  Hence, if the target condition is satisfied anywhere on the
negative-time branch, it is already satisfied at $t=0$ in the
one-sided problem.  Therefore the target condition is satisfied for
some $t\in\R$ under $\widetilde U$ if and only if it is satisfied for
some $t\geq0$ under $U$.

Thus, a total algorithm for the two-sided UHTP would give a total
algorithm for the one-sided UHTP.  This contradicts Theorem \ref{thm:fixed-uhtp}.
\end{proof}

The reduction can again be restricted to fixed dynamics.  Apply the
construction above to the single fixed evolution $U_*(t)$ of Theorem
\ref{thm:fixed-uhtp} and define
\[
\widetilde U_*(t)
=
\begin{cases}
U_*(t),
&
t\geq0,
\\
I,
&
t<0.
\end{cases}
\]
Then $\widetilde U_*(t)$ is a single fixed computably described
two-sided evolution, independent of $(M,x)$.  For the reduction
instances, the initial and target states are distinct computational
basis states, so the negative-time branch never satisfies the target
condition.  Consequently,
\[
M(x)\ {\rm halts}
\quad\Longleftrightarrow\quad
T_{\rm hit}^{\pm}
(\psi_{M,x},\phi_{M,x},\widetilde U_*,\varepsilon)
<
\infty.
\]
Thus the two-sided UHTP remains undecidable even when the two-sided
dynamics itself is fixed once and for all.

This result shows that the undecidability proved in this paper is not
an artifact of restricting the time parameter to $t\geq 0$.  Even when
both signs of the time parameter are allowed, the universal total
determination of the target hitting time is impossible in general.

\subsection{Relation to time-reversal symmetry}

The result is compatible with time-reversal symmetry of microscopic
laws.  What is ruled out is not time-reversal symmetry itself, but the
existence of a total universal procedure which selects, for all
inputs, a time at which a target state is reached.

The two-sided UHTP shows that, even when both orientations of the time
parameter are formally allowed, the universal time-selection problem
still contains the difficulty of the halting problem.  The claim is
therefore not a metaphysical statement about the arrow of time itself.
It is a statement about the nonexistence of a universal time-selection
procedure for computably described quantum dynamics.

The fixed-dynamics strengthening does not change this interpretation.
The two-sided construction above uses one and the same computably
described evolution for all encoded computations, but it is not
asserted to be time-reversal symmetric.  Its role is only to show that
neither allowing both signs of the time parameter nor fixing the
dynamical law removes the total-computability obstruction.

\section{Discussion and limitations}
\label{sec:discussion}

The result of this paper does not deny prediction or numerical
calculation for individual physical systems.  For a sufficiently simple
system, the hitting time may be analytically computable.  Reachability
or hitting times may also be decidable in restricted settings, such as
bounded time intervals, finite-dimensional systems with special
structure, integrable systems, or free systems.

What has been shown is that there is no total universal procedure for
the whole class of computably described unitary quantum dynamics.  More
strongly, the obstruction does not arise merely from allowing an
arbitrary input-dependent dynamical law.  Theorem
\ref{thm:fixed-uhtp} shows that there exists a single fixed computably
described evolution $U_*(t)$ for which the finite/infinite hitting-time
alternative is already undecidable when only the initial and target
computational-basis states are varied.

Thus the computational difficulty cannot be attributed simply to the
need to analyze an unrestricted family of Hamiltonians.  In the fixed
construction, the Hamiltonian schedule is bounded, piecewise constant,
periodic, and independent of the simulated Turing machine and its
input.  The unbounded computation is encoded in the state and is
carried forward by repeated application of the same dynamical law.

The result applies to classes of physical systems capable of embedding
universal computation.  Conversely, for restricted nonuniversal
classes, the hitting-time problem may remain decidable.  The no-go
theorem is therefore not a denial of individual calculations for all
physical systems, but a denial of a uniform total solver for a
universal class.

At the computability-theoretic level, the obstruction is also not
essentially quantum.  The core of the fixed-dynamics reduction is the
orbit of a single computable reversible bijection
\[
w_*:\Ccal\to\Ccal
\]
on classical configurations, for which reaching a designated beacon
configuration is equivalent to halting of the simulated computation.
The permutation unitary $W_*$ and its continuous-time realization
embed this reversible classical dynamics into the quantum formalism.
Accordingly, an analogous undecidability statement already holds for
the corresponding classical reversible reachability or hitting-time
problem.  Quantum mechanics supplies a natural unitary realization of
the construction, but the total-computability obstruction itself
originates in reversible universal computation and unbounded logical
time.

This observation also clarifies the role of time in the UHTP.  The essential unbounded resource is not a specifically quantum resource such as entanglement or interference, but the number of logical steps
that may have to occur before the target event is reached.  If
the embedded computation halts, the event occurs after a finite
physical time; if it does not halt, no finite target-reaching time
exists.  A total procedure must distinguish these two cases and hence
must solve the same finite/infinite alternative as the halting problem.

Variants involving noise, finite temperature, or finite-size effects
require care.  In the exact construction used here, halting and nonhalting instances
are separated by target fidelities $1$ and $0$, respectively.  This
finite separation may be useful in a robustness analysis, but it does
not by itself imply stability under perturbations over arbitrarily long
evolution times.  A quantitative robustness theorem would require
explicit error models, measurement models, and control of error
accumulation, and is left for future work.

Similarly, the operational consequence for finite-resource protocols
does not derive detailed thermodynamic lower bounds.  The claim is the
computability-theoretic no-go statement that, if a uniformly bounded
finite-time, finite-resource protocol could return $\Thit$ for all
inputs, then it would decide the halting problem.  The fixed-dynamics
construction strengthens this interpretation because the hypothetical
protocol need not implement an arbitrary input-dependent evolution:
the same fixed Hamiltonian schedule may be used for every encoded
instance.

Thus the result should be understood not as saying that hitting times
in quantum dynamics are generally impossible to study, but as saying
that there is no universal total procedure which returns the correct
hitting time for all inputs, even under a suitably chosen single fixed
dynamical law.

The two-sided formulation may also be viewed as a minimal analogue of
situations in foundational physics where neither a temporal orientation
nor a global time parameter is fixed in advance.  This observation is
only suggestive: the present paper does not address the full problem of
time in generally covariant theories.  It shows, rather, that even when
both signs of a time parameter are allowed, universal selection of a
target-reaching time remains obstructed by computability.

The result therefore suggests that computability provides a natural
interface between foundations of quantum theory and foundations of
mathematics.  The obstruction identified here is not peculiar to a
specific quantum model.  It arises from the coexistence of reversible
universal computation, unbounded evolution time, and the requirement
of a total decision procedure.  The UHTP provides a simple quantum
setting in which limitations of computation appear directly as
limitations on the universal determination of physical event times.

\section{Conclusion}
\label{sec:conclusion}

We have formulated a hitting-time problem for computably described
unitary quantum dynamics.  The hitting time is defined as the first
time at which the fidelity with a target state exceeds a prescribed
threshold, with value $\infty$ if the target is never reached.

The main result is that no total algorithm can decide, for all
computably described inputs, whether this hitting time is finite.
Consequently, no total algorithm can return the hitting time
\[
\Thit\in[0,\infty]
\]
for all inputs.

Under the broad class of computably described dynamics, this
undecidability already admits a direct reduction from the halting
problem in which the simulated computation is encoded into a
computable time-dependent schedule.  More strongly, however, we have
shown that such input-dependent dynamics are not necessary.  There
exists a single fixed computably described unitary evolution $U_*(t)$,
generated by a bounded piecewise-constant Hamiltonian that is periodic
with period $2$, for which the UHTP remains undecidable.

For every Turing machine $M$ and input $x$, only the initial and target
computational-basis states depend on $(M,x)$.  These states can be
constructed effectively so that
\[
M(x)\ {\rm halts}
\quad\Longleftrightarrow\quad
\exists t\geq0:
|\amp{\phi_{M,x}}{U_*(t)}{\psi_{M,x}}|^2
=
1.
\]
Thus the undecidability is not an artifact of allowing the external
time dependence itself to encode the simulated computation.  The
unbounded logical computation can instead be carried by the state
evolving under one and the same fixed dynamical law.

This fixed-dynamics result strengthens the connection between the
halting problem and the UHTP while preserving the central distinction
between Boolean reachability and total hitting-time computation.
Returning $\Thit$ requires not only finding a finite target-reaching
time when one exists, but also terminating and returning $\infty$ when
no such finite time exists.  It is this totality requirement that
produces the universal computability obstruction.

It follows that there is no universal total procedure for selecting a
target-reaching time.  Under the standard effectiveness assumption for
uniformly specified physical protocols, there is likewise no uniformly
bounded finite-time, finite-resource protocol that returns the hitting
time for all computably described inputs.

We have also introduced a two-sided version of the hitting-time problem
and shown that the same undecidability persists when both signs of the
time parameter are allowed.  Thus the no-go theorem is not an artifact
of restricting time to the positive direction, but a
computability-theoretic limitation on universal time selection itself.

The result does not deny the computability of hitting times for
particular solvable systems or restricted decidable classes.  What is
ruled out is a total procedure that works uniformly over the full class
of computably described quantum dynamics.  The UHTP therefore provides
a setting in which the limitations of universal computation appear
directly as limitations on the universal determination of physical
event times.



\end{document}